\documentclass[a4paper,12pt]{article}

\usepackage{amssymb}
\usepackage{mathrsfs}
\usepackage{bbm}
\usepackage{multirow}

\newlength{\dinwidth}
\newlength{\dinmargin}
\newcommand{\qsq}{q^2}

\begin{document}

\title{\bf Determination of the strong coupling $g_{B^* B\pi}$ from semi-leptonic $B\to \pi \ell \nu$ decay}\bigskip

\author{Xin-Qiang Li$^{1,2}$, Fang Su$^{3}$ and Ya-Dong Yang$^{3,4}$\\
{ $^1$\small Department of Physics, Henan Normal University, Xinxiang, Henan 453007, P.~R. China}\\
{ $^2$\small IFIC, Universitat de Val\`encia-CSIC, Apt. Correus 22085, E-46071 Val\`encia, Spain}\\
{ $^3$\small Institute of Particle Physics, Huazhong Normal University, Wuhan, Hubei  430079, P.~R. China}\\
{ $^4$\small Key Laboratory of Quark \& Lepton Physics, Ministry of Education, Huazhong Normal University}\\ 
     {\small Wuhan, Hubei, 430079, P.~R. China}}

\date{}
\maketitle
\bigskip\bigskip

\vspace{-1.5cm}

\begin{abstract}
{\noindent}
According to heavy-meson chiral perturbation theory, the vector form factor $f_+(q^2)$ of exclusive semi-leptonic decay $B\to \pi \ell \nu$ is closely related, at least in the soft-pion region (i.e., $q^{2} \sim (m_B-m_{\pi})^2$), to the strong coupling $g_{B^* B\pi}$ or the normalized coupling $\hat g$. Combining the precisely measured $q^2$ spectrum of $B\to \pi \ell \nu$ decay by the BaBar and Belle collaborations with several parametrizations of the form factor $f_+(q^2)$, we can extract these couplings from the residue of the form factor at the $B^*$ pole, which relies on an extrapolation of the form factor from the semi-leptonic region to the unphysical point $q^2=m_{B^*}^2$. Comparing the extracted values with the other experimental and theoretical estimates, we can test these various form-factor parametrizations, which differ from each other by the amount of physical information embedded in. It is found that the extracted values based on the BK, BZ and BCL parametrizations are consistent with each other and roughly in agreement with the other theoretical and lattice estimates, while the BGL ansatz, featured by a spurious, unwanted pole at the threshold of the cut, gives a neatly larger value.

\end{abstract}
\newpage

\section{Introduction}
\label{sec:intro}

The most promising decay mode for a precise determination of the Cabibbo-Kobayashi-Maskawa (CKM)~\cite{CKM} matrix element $|V_{ub}|$, both experimentally and theoretically, is the exclusive semi-leptonic $B \to \pi \ell \nu$ decay~\cite{Antonelli:2009ws}, for which a number of measurements have been made  by various collaborations~(CLEO~\cite{Athar:2003yg}, BaBar~\cite{:2010uj,delAmoSanchez:2010zd,Aubert:2006px} and Belle~\cite{Ha:2010rf,Hokuue:2006nr}). A fit to the measured $q^2$ spectrum, on the other hand, allows for a precise extraction of the $q^2$ dependence of the vector form factor $f_+(q^2)$, and thus provides a stringent check on our understanding of the dynamics of hadrons governed by QCD.

The heavy-to-light form factors are complicated nonperturbative objects, which have attracted extensive investigations in the literature. Besides various quark models~(see, e.g., \cite{ISGW2,quark-model}), which in many aspects help our phenomenological understanding of the heavy-to-light transitions, there exist two more quantitative predictions based on first principles of QCD, the lattice QCD~(LQCD) simulation~(see, e.g., \cite{Bailey:2008wp,Bernard:2009ke,LQCD-D-B}) and the QCD sum rules on the light-cone~(LCSR)~(see, e.g., \cite{Ball:2004ye,LCSR-B,LCSR-review,Huang}). These two methods are complementary to each other with respect to the momentum transfer $q^2$: while the LQCD calculations are restricted to the high $q^2$ region, reliable predictions of the LCSR method can only be made at the low $q^2$ region. 

Due to our limited theoretical knowledge of the $q^2$ dependence of the transition form factors, a variety of parametrizations have been proposed in the literature, trying to capture as much information as possible on the dynamics of the corresponding mesons. These include the two-parameter Be\'cirevi\'c-Kaidalov~(BK) ansatz~\cite{Becirevic:1999kt}, the three-parameter Ball-Zwicky~(BZ) ansatz~\cite{Ball:2004ye,Ball:2006jz}, the so-called Series Expansion~(SE) ansatz~\cite{Arnesen:2005ez,Becher:2005bg,BGL,BCL}, as well as the representation from the Omnes solution to the dispersive bounds~\cite{Omnes}. It turns out that most of them could fit the data equally well in the semi-leptonic region~\cite{:2010uj,Ha:2010rf,Ball:2006jz}. A good review of these different parametrizations could be found, for example, in Refs.~\cite{:2010uj,Ball:2006jz}.

Most of the above parametrizations include the essential feature that the vector form factor $f_+(q^2)$ has a pole at $q^2=m_{B^*}^2$, where $B^*(1^-)$ is a narrow resonance with $m_{B^*}=5.325~{\rm GeV}<m_B+m_{\pi}$. As the high-precision experimental data on $B\to\pi\ell\nu$ decay is available only in the semi-leptonic region, $0 \leq q^2 \leq (m_B-m_{\pi})^2$, in order to extract the pole residue we have to extrapolate the form factor from this region to the unphysical point $q^2=m_{B^*}^2$. Although lying outside the physical region, the pole residue is of great phenomenological interest. It is related to the strong coupling $g_{B^* B\pi}$, describing the low-energy interaction among the two heavy B-mesons and a soft pion, or the normalized coupling $\hat g$, a fundamental parameter in heavy-meson chiral perturbation theory~(HMChPT)~\cite{HMChPT,Casalbuoni:1996pg}. Since the process $B^* \to B \pi$ is kinematically forbidden, the coupling $g_{B^* B\pi}$ cannot be measured directly but should be fixed phenomenologically.
In this paper, exploiting the experimental knowledge on the form factor $f_+(q^2)$ extracted from the semi-leptonic $B\to\pi\ell\nu$ decay, we determine the strong coupling $g_{B^* B\pi}$ and $\hat g$ from the pole residue by extrapolating the form factor from the physical region to the unphysical point $q^2=m_{B^*}^2$. By comparing the extracted values with other theoretical and experimental estimates, we can then test the various form-factor parametrizations.

Our paper is organized as follows. In Section~\ref{sec:theo}, we provide the definition of heavy-to-light form factors, their different parametrizations, and the pole residue at $q^2=m_{B^*}^2$. In Section~\ref{sec:results}, after collecting the up-to-date measured $B\to \pi$ form-factor shape parameters, we give our determinations of the strong coupling $g_{B^* B\pi}$ and the corresponding normalized coupling $\hat g$; some interesting phenomenological discussions are also presented in this section. Our conclusions are made in Section~4.

\section{Heavy-to-light form factor}
\label{sec:theo}

\subsection{Definition of the heavy-to-light form factor}
\label{sec:FF-def}

In exclusive semi-leptonic $B\to\pi\ell\nu$ decay, the hadronic matrix element is usually parameterized in terms of two form factors $f_+(q^2)$ and $f_0(q^2)$~\cite{Wirbel:1985ji},
\begin{equation}\label{eq:ffdef1}
  \langle \pi(p_{\pi}) | \bar u \gamma^{\mu} b| \bar{B}(p_B) \rangle  = f_+(q^2) \left[(p_B + p_{\pi})^{\mu} - \frac{m_B^2 - m_{\pi}^2}{q^2}\, q^\mu \right] + f_0(q^2) \frac{m_B^2 - m_{\pi}^2}{q^2}\,q^\mu\,,
\end{equation}
where $q \equiv p_B - p_{\pi}$ is the momentum transferred to the lepton pair, with $p_B$ and $p_{\pi}$ the four-momenta of the parent B-meson and the final-state pion, and $m_B$ and $m_{\pi}$ their masses. For massless leptons, which is a good approximation for electrons and muons, the form factor $f_0(q^2)$ is absent and we are left with only a single form factor $f_+(q^2)$.

Precise knowledge of the heavy-to-light form factors is of primary importance for flavour physics. It is needed for the determination of the CKM matrix element $|V_{ub}|$ from exclusive semi-leptonic $B\to\pi\ell\nu$ decay. They are also needed as ingredients in the analysis of hadronic B-meson decays, such as $B\to \pi\pi$ and $B \to \pi K$, in the framework of QCD factorization~\cite{Beneke:1999br}, again with the objective to provide precision determinations of the quark flavour mixing parameters.

The two QCD methods, LQCD and LCSR, result in predictions for different $q^2$ regions. The LCSR combines the idea of QCD sum rules with twist expansions performed up to ${\cal O}(\alpha_s)$, and provides estimates of various form factors at low intermediate $q^2$ regions, $0<q^2<14~{\rm GeV}^2$. The overall normalization is predicted at the zero momentum transfer with typical uncertainties of $10-13\%$~\cite{Ball:2004ye,LCSR-B}. The LQCD simulation can, on the other hand, potentially provide the heavy-to-light form factors in the high-$q^2$ region from first principles of QCD. The unquenched lattice calculations, in which quark-loop effects in the QCD vacuum and three dynamical quark flavours~(the mass-degenerate $u$ and $d$ quarks and a heavier $s$ quark) are incorporated, are now available for $B\to \pi$ form factors~\cite{Bailey:2008wp,Bernard:2009ke,LQCD-D-B}. Unfortunately, neither the LQCD nor the LCSR can predict the form factors over the full $q^2$ range.

\subsection{Form-factor parametrizations}
\label{sec:FF-para}

While predictions of the exact form-factor shape are challenged for any theoretical calculations, it is well established that the general properties of analyticity, crossing symmetry and unitarity largely constrain the $q^2$ behavior of the form factor~\cite{Becher:2005bg,BGL,BCL}. Specifically, it is expected to be an analytic function everywhere in the complex $q^2$ plane outside of a cut that extends along the positive $q^2$ axis from the mass of the lowest-lying $b \bar{d}$ vector meson. This assumption leads to an un-subtracted dispersion relation~\cite{Becher:2005bg},
\begin{equation}\label{eq:dispersion}
  f_+(q^2) = \frac{f_+(0)/(1-\alpha)}{1-\qsq/m^2_{B^*}} + \frac{1}{\pi}\int_{(m_B+m_{\pi})^2}^\infty dt {\frac{{\rm Im}f_+(t)}{t-\qsq-i\epsilon}}\,,
\end{equation}
which means that we have a pole residue at $q^2=m_{B^*}^2$ and a cut from the $B\,\pi$ continuum, and the parameter $\alpha$ gives the relative size of contribution to $f_+(0)$ from the $B^*$ pole.

The various parametrizations proposed in the literature make explicitly or implicitly different simplifications in the treatment of the cut, and the following four ones are widely used, with their respective salient features sketched below:
\begin{enumerate}
\item{Be\'cirevi\'c-Kaidalov~(BK) ansatz~\cite{Becirevic:1999kt}:}
\begin{equation} \label{eq:BK-ansatz}
  f_+(q^2) = \frac{f_+(0)} {(1-q^2/m_{B^*}^2)(1-\alpha_{BK}\, q^2/m_{B^*}^2)}\,,
\end{equation}
where $f_+(0)$ sets the normalization and $\alpha_{BK}$ defines the shape of the form factor. It is mainly motivated by the scaling laws of the form factors in the heavy quark limit, and provides an approximate representation of the second term in Eq.~(\ref{eq:dispersion}) by an additional effective pole $m_{B^*}^2/\alpha_{BK}$, with $\alpha_{BK}<1$ to be consistent with the location of the cut.

\item{Ball-Zwicky~(BZ) ansatz~\cite{Ball:2004ye,Ball:2006jz}:}
\begin{equation} \label{eq:BZ-ansatz}
  f_+(q^2) = f_+(0) \left[ \frac{1} {1-q^2/m_{B^*}^2} +
  \frac{r_{BZ}\, q^2/m^2_{B^*}}{(1-q^2/m_{B^*}^2)\,(1-\alpha_{BZ}\, q^2/m_{B^*}^2)} \right]\,,
\end{equation}
where $f_+(0)$ is the normalization, and $\alpha_{BZ}$ and $r_{BZ}$ determine the shape of the form factor. This is an extension of the BK ansatz, related to each other by the simplification $\alpha_{BK}=\alpha_{BZ}=r_{BZ}$. The BK and BZ parametrizations are featured by both being intuitive and having fewer free parameters.

\item{Boyd-Grinstein-Lebed~(BGL) ansatz~\cite{Becher:2005bg,BGL}:}
\begin{equation} \label{eq:BGL-ansatz}
 f_+(q^2) = \frac{1}{P(q^2) \phi(q^2,q^2_0)}\, \sum_{k=0}^{k_{max}} a_k(q^2_0) \big[z(q^2,q^2_0)\big]^k \,,
\end{equation}
with the conformal mapping variable defined by
\begin{equation} \label{eq:z-def}
 z(q^2,q^2_0) = \frac {\sqrt{t_+ - q^2} - \sqrt{t_+ - q^2_0}}{\sqrt{t_+ - q^2} + \sqrt{t_+ - q^2_0}}\,,
\end{equation}
where $t_{\pm} = (m_B \pm m_{\pi})^2$ and $q^2_0$ is a free parameter. The so-called Blaschke factor $P(q^2)=z(q^2,m_{B^*}^2)$ accounts for the pole at $q^2=m_{B^*}^2$, and the outer function $\phi(q^2,q^2_0)$ is an arbitrary analytic function, the choice of which affects only the particular values of the series coefficients $a_k$. The form-factor shape is determined by the values of $a_k$, with truncation at $k_{max}=2$ or $3$. The expansion parameters $a_k$ are bounded by unitarity, $\sum_k a_k^2 \le 1$. Becher and Hill~\cite{Becher:2005bg} have pointed out that due to the large $b$-quark mass, this bound is far from being saturated. For more details we refer to Refs.~\cite{Becher:2005bg,BGL}.

\item{Bourrely-Caprini-Lellouch~(BCL) ansatz~\cite{BCL}:}
\begin{equation} \label{eq:BCL-ansatz}
 f_+(q^2) = \frac{1} {1-q^2/m_{B^*}^2} \sum_{k=0}^{k_{max}} b_k \left\{[z(q^2,q^2_0)]^k - (-1)^{k-k_{max}-1} \frac{k}{k_{max}+1} [z(q^2,q^2_0)]^{k_{max}+1}\right\} \,,
\end{equation}
where the variable $z(q^2,q^2_0)$ is defined by Eq.~(\ref{eq:z-def}), and the free parameter $q^2_0$ can be chosen to make the maximum value of $|z|$ as small as possible in the semi-leptonic region~\cite{BCL}. In this ansatz, the form-factor shape is determined by the values of $b_k$, with truncation at $k_{max}=2$ or $3$.
\end{enumerate}

Although the BK and the BZ parametrization are intuitive and have few free parameters, the presence of poles near the semi-leptonic region creates doubt on whether truncating all but the first one or two terms leaves an accurate estimate of the true form-factor shape. The BGL and the BCL parametrization are based on some fundamental theoretical concepts like analyticity and unitarity, and avoid \textit{ad hoc} assumptions about the number of poles and the pole masses. Fits to the measured $q^2$ spectrum of $B\to \pi \ell \nu$ decay have, on the other hand, shown that these different form-factor parametrizations could describe the data equally well~\cite{:2010uj}.

\subsection{Pole residue at $q^2=m_{B^*}^2$ and the strong coupling $g_{B^* B\pi}$}
\label{sec:FF-residue}

All the above four parametrizations have the essential feature that the vector form factor $f_+(q^2)$ has a pole at $q^2=m_{B^*}^2$. Although lying outside the semi-leptonic region, the pole residue at $q^2=m_{B^*}^2$ is phenomenologically very interesting. With the following standard definitions~\cite{Becirevic:1999kt},
\begin{equation} \label{eq:decay-def}
\langle 0|\bar{d}\gamma_{\mu}b|\bar{B}^{*0}(p,\epsilon) \rangle = f_{B^*}m_{B^*}\epsilon_{\mu}, \qquad
\langle B^-(p)\pi^+(q)|\bar{B}^{*0}(p+q,\epsilon) \rangle = g_{B^* B\pi}(q\cdot \epsilon)\,,
\end{equation}
it is given by the product of the strong coupling $g_{B^* B\pi}$ and the vector decay constant $f_{B^*}$~\cite{Ball:2004ye,Becirevic:1999kt},
\begin{eqnarray}\label{eq:r1-def}
 r_1 &=& \lim_{\qsq=m_{B^*}^2} (1-\qsq/m_{B^*}^2)\,f_+(\qsq)\, \nonumber \\
     &=& \frac{f_{B^*}\,g_{B^* B\pi}}{2 m_{B^*}}\,.
\end{eqnarray}
In fact, at the upper end of the physical region~(i.e., at the zero recoil point $q^2=(m_B-m_{\pi})^2$), the vector-meson dominance~(VMD) of $f_+(q^2)$ is expected to be very effective~\cite{Isgur:1989qw,Burdman:1993es}. It has been argued that, in the combined heavy quark and chiral limit, the VMD becomes even exact~\cite{Grinstein:1994nx}. Thus, the strong coupling $g_{B^* B\pi}$ determines the normalization of the vector form factor $f_+(q^2)$ near the zero recoil of pion. The strong coupling $g_{B^* B\pi}$ also provides access to the normalized coupling $\hat g$, which is, in the limit of exact chiral, heavy flavour and spin symmetries, the single parameter for heavy-meson chiral perturbation theory~(HMChPT)~\cite{HMChPT,Casalbuoni:1996pg}. They are related to each other through~\cite{Abada:2002xe}
\begin{equation} \label{eq:ghat-def}
 \hat g = \frac{g_{B^* B\pi}}{2\,\sqrt{m_B m_{B^*}}}\,f_{\pi}\,,
\end{equation}
where the convention $f_{\pi} \simeq 131~{\rm MeV}$ is used. Unlike the $D^* D\pi$ coupling $g_{D^* D\pi}$, which could be extracted from the available experimental data on the decay $D^* \to D\pi$~\cite{Ahmed:2001xc}, there cannot be a direct experimental indication on the coupling $g_{B^* B\pi}$, because there is no phase space for the decay $B^* \to B\pi$. They could however be related through the heavy quark symmetry~\cite{Casalbuoni:1996pg}.

As a result, a precise determination of the couplings $g_{B^* B\pi}$ and $\hat g$ is of particular importance. During recent years a large number of theoretical studies have been devoted to the calculation of these couplings in various versions of quark models~\cite{Isgur:1989qw,quarkmodel} and QCD sum rules~\cite{qcdsr1,qcdsr2}. However, the variation of the obtained values, even within a single class of models, turns out to be quite large~\cite{LCSR-review,Casalbuoni:1996pg}, for an overview see~\cite{LCSR-review,Casalbuoni:1996pg}\footnote{Values for the couplings obtained prior to 1995 with different approaches could be found, for example, in \cite{qcdsr2} and references therein.}. In addition, there have been several LQCD simulations of these couplings in both quenched~\cite{deDivitiis:1998kj,Abada:2003un} and unquenched~\cite{Becirevic:2009yb,Ohki:2008py} approximations. These strong couplings have also been calculated using a framework based on QCD Dyson-Schwinger equations~\cite{ElBennich:2010ha,Ivanov:1998ms}.

Motivated by the precise experimental knowledge on the vector form factor $f_+(q^2)$, one can extract indirectly the values of $g_{B^* B\pi}$ via Eq.~(\ref{eq:r1-def}) and $\hat g$ via Eq.~(\ref{eq:ghat-def}), by an extrapolation of the form factor from the physical region to the pole $m_{B^*}^2$, which will be detailed in the next section.

\section{Numerical results and discussions}
\label{sec:results}

\subsection{The fitted $B\to \pi$ form-factor shape parameters}
\label{sec:FF-shape-paras}

In order to extrapolate the vector form factor $f_+(q^2)$ to the $B^*$ pole based on the various form-factor parametrizations, we first need to determine their shape parameters from the current experimental data on $B\to \pi \ell \nu$ decay reported by the BaBar~\cite{:2010uj,delAmoSanchez:2010zd,Aubert:2006px} and  Belle~\cite{Ha:2010rf,Hokuue:2006nr} collaborations. Although these measurements employ different experimental techniques in treating the second B meson in the $B\bar{B}$ event, the measured total and partial branching fractions agree well among each other. For more details, we refer to these original references~\cite{:2010uj,delAmoSanchez:2010zd,Aubert:2006px,Ha:2010rf,Hokuue:2006nr}.

These experiments have also measured the $q^2$ spectrum of $B\to \pi \ell \nu$ decay, a fit to which allows for an extraction of the $q^2$ dependence of the vector form factor $f_+(q^2)$. It is generally observed that all the four form-factor parametrizations introduced in section~\ref{sec:FF-para} could describe the measured spectrum equally well~\cite{:2010uj,Ha:2010rf,Ball:2006jz}. A summary of the fitted form-factor shape parameters based on various parametrizations is given in Table~\ref{tab:FFBtopisumary}, where both a linear~(2 para., with $k_{max}=2$) and a quadratic~(3 para., with $k_{max}=3$) ansatz for the BGL and BCL parametrizations are considered in \cite{:2010uj}, while a third-order polynomial fit~(4 para., with $k_{max}=4$) is perfermed in \cite{Ha:2010rf}. The value of the product $|V_{ub}|f_+(0)$ obtained from the fit extrapolated to $q^2=0$, if available, are listed in the last column. 

\begin{table}[t]
\begin{center}
\caption{\label{tab:FFBtopisumary} \small Summary of the form-factor shape parameters obtained by fitting to the BaBar~\cite{:2010uj}~(top) and Belle~\cite{Ha:2010rf}~(bottom) measurements for the isospin-combined $B\to \pi \ell \nu$ decays, based on various parametrizations of the vector form factor $f_+(q^2)$. }
\vspace{0.2cm}
\doublerulesep 0.8pt \tabcolsep 0.35in
\begin{tabular}{lll} \hline\hline
Parametrization & Fit parameters & $|V_{ub}|f_+(0)~[10^{-3}]$ \\
\hline
BK           & $\alpha_{BK} =+0.310\pm 0.085$       & $1.052 \pm 0.042$~\cite{:2010uj} \\[0.2cm]
BZ           & $r_{BZ} =+0.170 \pm 0.124$           & $1.079  \pm 0.046$~\cite{:2010uj} \\
             & $\alpha_{BZ} =+0.761 \pm 0.337$      &                                   \\[0.2cm]
BCL~(2 par.) & $b_1/b_0 = -0.67 \pm 0.18$    & $1.065 \pm 0.042$~\cite{:2010uj} \\[0.2cm]
BCL~(3 par.) & $b_1/b_0 = -0.90 \pm 0.46$    & $1.086 \pm 0.055$~\cite{:2010uj} \\
             & $b_2/b_0 = +0.47 \pm 1.49$    &                                  \\
BGL~(2 par.) & $a_1/a_0=-0.94 \pm 0.20$      & $1.103 \pm 0.042$~\cite{:2010uj} \\[0.2cm]
BGL~(3 par.) & $a_1/a_0=-0.82 \pm 0.29$      & $1.080 \pm 0.056$~\cite{:2010uj} \\
             & $a_2/a_0=-1.14 \pm 1.81$      &                                  \\
\hline
BK           & $\alpha_{BK} =+0.60 \pm 0.04 $ & $0.924 \pm 0.028 $~\cite{Ha:2010rf} \\[0.2cm]
BGL~(4 par.) & $a_0=+0.022 \pm 0.002$         & $---             $~\cite{Ha:2010rf} \\
             & $a_1=-0.032 \pm 0.004$         &                   \\
             & $a_2=-0.080 \pm 0.020$         &                   \\
             & $a_3=+0.081 \pm 0.066$         &                   \\
\hline \hline
\end{tabular}
\end{center}
\end{table}

As concluded in Refs.~\cite{:2010uj,Ball:2006jz}, all these form-factor parametrizations could describe the experimental data equally well, and the central values of the product $|V_{ub}|f_+(0)$ agree with each other. Thus, all the four form-factor parametrizations are valid choices to describe the $q^2$ dependence of the vector form factor $f_+(q^2)$, at least in the physical region. To further test these different form-factor parametrizations, more precise and additional information is needed.

\subsection{The relevant input parameters}
\label{sec:input-paras}

Before presenting the results for the strong coupling $g_{B^* B\pi}$, we would like to first fix the relevant input parameters, such as the decay constants, the CKM matrix element $|V_{ub}|$, as well as the free parameter $q^2_0$ in the BGL and BCL parametrizations.

The vector decay constant defined by Eq.~(\ref{eq:decay-def}) is not relevant from a phenomenological point of view, since the meson $B^*$ will decay predominantly through the electromagnetic interaction. It is, however, needed in our case to extract the strong coupling $g_{B^* B\pi}$ from the pole residue Eq.~(\ref{eq:r1-def}). To take into account the uncertainties induced by this quantity, we shall use the following two inputs: one is taken from the UKQCD collaboration~\cite{Bowler:2000xw},
\begin{equation} \label{eq:fBstar-value1}
\tilde{f}_{B^*}=28(1)^{+3}_{-4}\,,
\end{equation}
which is related to the vector decay constant by $\tilde{f}_{B^*}=m_{B^*}/f_{B^*}$, with the first error quoted statistical and the second systematic, and hence we get $f_{B^*}=(190 \pm 7_{stat.}{}{^{+32}_{-18}}_{syst.})~{\rm MeV}$; the other one is taken from the quenched LQCD calculation~\cite{Bernard:2001fz},
\begin{equation} \label{eq:fBstar-value2}
f_{B^*}=(177 \pm 6_{stat.} \pm 17_{syst.})~{\rm MeV}\,.
\end{equation}

To extract the normalized form factor $f_+(0)$ from the fitted results of the product $|V_{ub}|f_+(0)$, one needs to know the value of the CKM matrix element $|V_{ub}|$. The two avenues for $|V_{ub}|$ determination through inclusive and exclusive $b \to u \ell \nu$ decays have been reviewed in~\cite{HFAG,PDG10}. How to reconcile the difference between the values for $|V_{ub}|$ obtained from these two methods remains an intriguing puzzle. At the same time, $|V_{ub}|$ can also be most precisely determined by a global fit of the unitarity triangle~(UT) that uses all available measurements~\cite{CKMfit,UTfit}. Since the presence of New Physics~(NP) might, in principle, affect the result of the UT analysis, here we shall use the tree-level fit result performed by the UTfit collaboration~\cite{UTfit}, 
\begin{equation} \label{eq:Vub-input}
|V_{ub}|=(3.76 \pm 0.20) \times 10^{-3}\,,
\end{equation}
which is almost unchanged by the presence of NP.

In the BGL and BCL parametrizations, both the free parameter $q_0^2$ and the outer function $\phi(q^2,q_0^2)$ have to be specified. Following the BaBar collaboration~\cite{:2010uj} and references therein, we choose the values $q_0^2=0.65 t_-$ for the BGL, and $q_0^2=(m_B+m_{\pi})(\sqrt{m_B}-\sqrt{m_{\pi}})^2$ for the BCL parametrization. The outer function $\phi(q^2,q_0^2)$ in the BGL parametrization is given explicitly as~\cite{Arnesen:2005ez},
\begin{eqnarray}  \label{eq:slep:phi}
    \phi_+(q^2, q_0^2)  &=&  \sqrt{\frac{1}{32 \pi \chi^{(0)}_J}}
    \big( \sqrt{t_+ - q^2} + \sqrt{t_+ - q_0^2}  \big)
    \big( \sqrt{t_+ - q^2} + \sqrt{t_+ - t_-} \big)^{3/2}  \nonumber \\
    & \times & \big( \sqrt{t_+ - q^2} + \sqrt{t_+} \big)^{-5}
    \frac{(t_+ - q^2)}{(t_+ - q_0^2)^{1/4}}\,,
\end{eqnarray}
where $\chi^{(0)}_J$ is a numerical factor that can be calculated via operator product expansion~\cite{Lellouch:1995yv}. 
At two loops in terms of the pole mass and condensates and taking $\mu=m_b$, it is given as~\cite{Arnesen:2005ez}
\begin{equation} \label{chiJ0-function}
 \chi^{(0)}_J = \frac{3\big[ 1\! +\! 1.140\, \alpha_s(m_b)\big]}{32\pi^2m_b^2} 
  \!-\! \frac{\overline{m}_b\:\langle  \bar u u\rangle}{m_b^6}
  \!-\! \frac{\langle \alpha_s G^2\rangle}{12\pi m_b^6} \,,
\end{equation}
with $m_b=4.88~{\rm GeV}$, $\overline{m}_b \langle \bar u u\rangle \simeq -0.076\,{\rm GeV}^4$, $\langle \alpha_s G^2 \rangle \simeq 0.063 {\rm GeV}^4$~\cite{Arnesen:2005ez}. Explicitly the BaBar collaboration uses $\chi^{(0)}_J=6.889 \times 10^{-4}$~\cite{:2010uj}.

For all the other input parameters, we list them in Table~\ref{tab:Inputs}. Throughout the paper, we use the isospin-averaged meson masses, for example, $m_{\pi}=(m_{\pi^+}+m_{\pi^0})/2$. 

\begin{table}[t]
\begin{center}
\caption{\label{tab:Inputs} \small The relevant input parameters used in our calculation. All meson masses are taken directly from the Particle Data Group~\cite{PDG10}.}
\vspace{0.2cm}
\doublerulesep 0.8pt \tabcolsep 0.15in
\begin{tabular}{lll}
\hline \hline
$m_{\pi^+}=139.6~{\rm MeV}$ & $m_{\pi^0}=135.0~{\rm MeV}$ &
$f_{\pi}=130.41 \pm 0.20~{\rm MeV}$~\cite{PDG10} \\
$m_{B^+}=5279.2~{\rm MeV} $ & $m_{B^0}=5279.5~{\rm MeV} $ & $m_{B^{\ast}}=5325.1~{\rm MeV}$ \\
\hline \hline
\end{tabular}
\end{center}
\end{table}

\subsection{Numerical results for the couplings $g_{B^* B\pi}$ and $\hat g$}

In this subsection, assuming a definite behavior of the $q^2$ dependence of the vector form factor $f_+(q^2)$ and using the fitted shape parameters listed in Table~\ref{tab:FFBtopisumary}, we shall extrapolate the form factor to the $B^*$ pole and extract the strong couplings $g_{B^* B\pi}$ and $\hat g$ through Eqs.~(\ref{eq:r1-def}) and (\ref{eq:ghat-def}).

\subsubsection{The coupling  \boldmath $g_{B^* B\pi}$}

As mentioned already, 
the coupling $g_{B^* B\pi}$ is only poorly known phenomenologically and the literature exhibits a wide spread of values~\cite{LCSR-review,Casalbuoni:1996pg,deDivitiis:1998kj,Abada:2003un,Becirevic:2009yb,Ohki:2008py}. In this subsection, we first present in Table~\ref{tab:strongcoupling} the extracted values of $g_{B^* B\pi}$ from the pole residue.

\begin{table}[t]
\begin{center}
\caption{\label{tab:strongcoupling} \small The extracted values of the strong couplings $g_{B^* B\pi}$ and $\hat g$ using different form-factor parametrizations with the shape parameters given in Table~\ref{tab:FFBtopisumary}. The columns Eq.~(\ref{eq:fBstar-value1}) and Eq.~(\ref{eq:fBstar-value2}) denote the results obtained with the corresponding input for $f_{B^*}$ given by these two equations.}
\vspace{0.2cm}
\doublerulesep 0.8pt \tabcolsep 0.11in
\begin{tabular}{lccccc} \hline\hline
\multirow{2}{*}{Parametrization}   & \multirow{2}{*}{$f_{B^*}g_{B^* B\pi}~[{\rm GeV}]$} &
\multicolumn{2}{c}{$g_{B^* B\pi}$} & \multicolumn{2}{c}{$\hat g$} \\ \cline{3-6}
&                       & Eq.~(\ref{eq:fBstar-value1}) & Eq.~(\ref{eq:fBstar-value2})
                        & Eq.~(\ref{eq:fBstar-value1}) & Eq.~(\ref{eq:fBstar-value2}) \\
\hline
BK     & $4.32^{+0.68}_{-0.55}$  & $22.71^{+4.38}_{-4.42}$ & $24.40^{+4.72}_{-3.84}$
       & $0.28^{+0.05}_{-0.05}$  & $ 0.30^{+0.06}_{-0.05}$~\cite{:2010uj}     \\[0.2cm]
BZ     & $5.23^{+1.63}_{-1.62}$  & $27.50^{+9.11}_{-9.45}$ & $29.55^{+9.79}_{-9.57}$
       & $0.34^{+0.11}_{-0.12}$  & $ 0.36^{+0.12}_{-0.12}$~\cite{:2010uj} \\[0.2cm]
BCL~(2 par.)   & $ 4.82^{+0.74}_{-0.65}$ & $25.34^{+ 4.85}_{- 5.07}$ & $27.23^{+ 5.21}_{- 4.46}$
               & $ 0.31^{+0.06}_{-0.06}$ & $ 0.33^{+ 0.06}_{- 0.05}$~\cite{:2010uj} \\[0.2cm]
BCL~(3 par.)   & $ 5.78^{+2.11}_{-1.56}$ & $30.38^{+11.61}_{- 9.32}$ & $32.64^{+12.48}_{- 9.29}$
               & $ 0.37^{+0.14}_{-0.11}$ & $ 0.40^{+ 0.15}_{- 0.11}$~\cite{:2010uj} \\[0.2cm]
BGL~(2 par.)   & $10.57^{+1.60}_{-1.44}$ & $55.58^{+10.48}_{-11.16}$ & $59.72^{+11.28}_{- 9.85}$
               & $ 0.68^{+0.13}_{-0.14}$ & $ 0.73^{+ 0.14}_{- 0.12}$~\cite{:2010uj} \\[0.2cm]
BGL~(3 par.)   & $ 7.76^{+3.44}_{-3.89}$ & $40.81^{+18.67}_{-21.30}$ & $43.85^{+20.06}_{-22.33}$
               & $ 0.50^{+0.23}_{-0.26}$ & $ 0.54^{+ 0.25}_{- 0.27}$~\cite{:2010uj} \\
\hline
BK             & $ 6.54^{+0.77}_{-0.66}$  & $34.40^{+5.61}_{-6.15}$ & $36.97^{+6.04}_{-5.07}$
               & $ 0.42^{+0.07}_{-0.08}$  & $ 0.45^{+0.07}_{-0.06}$~\cite{Ha:2010rf} \\[0.2cm]
BGL~(4 par.)   & $ 0.34^{+4.59}_{-4.59}$ & $ 1.78^{+24.12}_{-24.12}$ & $1.92^{+25.91}_{-25.91}$
               & $ 0.02^{+0.30}_{-0.30}$ & $ 0.02^{+ 0.32}_{- 0.32}$~\cite{Ha:2010rf} \\
\hline \hline
\end{tabular}
\end{center}
\end{table}

Since the vector decay constant $f_{B^*}$ could not be measured directly and the lattice calculation still has a large uncertainty~\cite{Bowler:2000xw,Bernard:2001fz}, we also give the values of the product $f_{B^*}g_{B^* B\pi}$ in Table~\ref{tab:strongcoupling}, which is free of the uncertainty induced by $f_{B^*}$. Comparing the values listed in the two columns Eq.~(\ref{eq:fBstar-value1}) and Eq.~(\ref{eq:fBstar-value2}), we can see that the extracted values of $g_{B^* B\pi}$ and $\hat g$ are not so sensitive to the vector decay constant, and are consistent with each other within their respective error bars. Further reduction of the uncertainty on the vector decay constant $f_{B^*}$ is welcome from the LQCD simulation.

As can be seen from the upper part in Table~\ref{tab:strongcoupling}, the extracted results of the parameters based on all the four parametrizations are roughly consistent with each other with their respective uncertainties taken into account; the central values obtained with the BGL parametrization, on the other hand, are neatly larger than the ones with the other three parametrizations. As noted in Refs.~\cite{BCL,DescotesGenon:2008hh}, this is due to the spurious zero at $q^2=t_+$ in definition of the outer function $\phi(q^2,q^0)$ in Eq.~(\ref{eq:slep:phi}), implying that the BGL parametrization includes {\em a spurious, unwanted pole at the threshold of the cut}. Although being also a series-expansion-based ansatz, the BCL parametrization could yield a value in good agreement with the BK and BZ ones, which confirms the reason for generating such a larger value in the BGL ansatz caused by the spurious zero in $\phi(q^2,q^0)$. In addition, comparing the linear and the quadratic fits in the BGL and BCL parametrizations, we can see that the errors increase with more expansion parameters added, leading to a loss of predictive power. This means that the BGL and BCL parametrizations with more fitting parameters could not be well constrained  by the current data of  the semi-leptonic B decays.

From the lower part in Table~\ref{tab:strongcoupling}, on the other hand, we can see that, while the results of the BK parametrization are roughly consisitent with the ones using the other ansatz based the BaBar data~\cite{:2010uj}, the BGL parametrization performed by the Belle collaboration~\cite{Ha:2010rf} gives much smaller results, but with larger uncertainties. This might be due to the fact that the Belle collaboration~\cite{Ha:2010rf} uses a different fitting strategy: rather than treating the model-independent quantity $|V_{ub}|f_+(0)$ as a free parameter~(as does the BaBar collabortion~\cite{:2010uj}), they perform a simultaneous fit of the experimental~\cite{Ha:2010rf} and the FNAL/MILC~\cite{Bailey:2008wp} LQCD results, where the free parameters are the CKM matrix element $|V_{ub}|$ and the series-expansion parameters $a_i$. In order to compare directly with the BaBar results, a similar fit from the Belle collaboration is necessarily needed.

To check the validity of the form-factor extrapolation, we would like to compare the values of $f_{B^*}g_{B^* B\pi}$ given in Table~\ref{tab:strongcoupling} with the ones existing in the literature,
\begin{equation} \label{eq:fbgcoupling-ref}
f_{B^*} g_{B^* B\pi}=\left\{
\begin{array}{l}
(4.44 \pm 0.97)~{\rm GeV}~\cite{qcdsr2}\,, \\
(7.77,7.88,8.20,10.01)~{\rm GeV} \quad \mbox{for~sets~1~to~4}~\cite{Ball:2004ye}\,,
\end{array} \right.
\end{equation}
from which we can see that our results are generally consistent with them. On the other hand, it is observed that the result obtained in the LCSR method~\cite{qcdsr2} is smaller than the fits given in Ref.~\cite{Ball:2004ye}; this might be due to the failure of the simple quark-hadron duality used for the contribution of higher resonances and the continuum to the sum rules~\cite{Becirevic:2002vp}; the inclusion of a radial excitation with negative residue in the hadronic parametrization of the correlation function does increase the value~\cite{Becirevic:2002vp}. With this fact taken into account, our central values are a bit smaller than that given in Eq.~(\ref{eq:fbgcoupling-ref}).

\subsubsection{The normalized coupling \boldmath $\hat g$}

The normalized coupling $\hat g$ is the single constant in the limit of exact chiral, heavy flavour and spin symmetries~\cite{HMChPT,Casalbuoni:1996pg}. However, being the parameter of the effective theory, its value cannot be predicted but should be fixed phenomenologically. Our results are given in last two columns in Table~\ref{tab:strongcoupling}. As is the case for $g_{B^* B\pi}$, the central values based on the BK, BZ and BCL parametrizations are consistent with each other, while the ones in the BGL ansatz are larger.

As an improved determination of the $B^* B\pi$ coupling can reduce the systematic uncertainty in most lattice calculations of B-meson quantities, it has aroused a lot of precise determinations of the $B^* B\pi$ coupling in the literature~\cite{deDivitiis:1998kj,Abada:2003un,Becirevic:2009yb,Ohki:2008py}. The most recent lattice results are
\begin{equation} \label{eq:ghat-lattice}
\hat g=\left\{\begin{array}{l}
0.42 \pm 0.04_{stat} \pm 0.08_{syst} \qquad  {\rm for} N_f=0~\cite{deDivitiis:1998kj}\,, \\
0.58 \pm 0.06_{stat} \pm 0.10_{syst} \qquad  {\rm for} N_f=0~\cite{Abada:2003un}\,, \\
0.44 \pm 0.03_{stat} {}^{+0.07}_{-0.00}{}_{syst} \qquad {\rm for} N_f=2~\cite{Becirevic:2009yb}\,,\\
0.516 \pm 0.005_{stat} \pm 0.033_{chiral} \pm 0.028_{pert} \pm 0.028_{dics} \qquad {\rm for} N_f=2~\cite{Ohki:2008py}\,,
\end{array} \right.
\end{equation}
which have about $5\%$ and $15\%$ statistical errors for the quenched and unquenched cases, respectively. With their respective uncertainties taken into account, our extracted values are generally consistent with the above lattice data.

Other estimates of the coupling $\hat g$ are derived using various versions of quark models and QCD sum rules~\cite{LCSR-review,Casalbuoni:1996pg}. The best estimate based on the analyses of both QCD sum rules and relativistic quark model, quoted in the review~\cite{Casalbuoni:1996pg}, is
\begin{equation}
\hat g \simeq 0.38\,,
\end{equation}
with an uncertainty around $20\%$, which is also in agreement with our results given in Table~\ref{tab:strongcoupling}.

Both the strong couplings $g_{B^* B\pi}$ and $\hat g$ have also been calculated using a framework based on QCD's Dyson-Schwinger equations~\cite{ElBennich:2010ha,Ivanov:1998ms}. By implementing a more realistic representation of heavy-light mesons, the updated analysis based on this framework gives $g_{B^* B\pi}=30.0^{+3.2}_{-1.4}$ and $\hat g=0.37^{+0.04}_{-0.02}$~\cite{ElBennich:2010ha}, both of which are also consistent with our extracted values from the semi-leptonic $B\to \pi \ell \nu$ decays.

The coupling $\hat g$ is also related to the measured decay width $\Gamma(D^* \to D\pi)$~\cite{Ahmed:2001xc}. From the width of the charged $D^*$-meson measured by CLEO, $\Gamma^{\rm exp}(D^{*+})=(96 \pm 22)~{\rm KeV}$~\cite{Ahmed:2001xc}, and by using the experimentally established branching fraction ${\mathcal B} (D^{*+}\to D^+ \gamma)=(1.6 \pm 0.4)\%$~\cite{PDG10}, we can get
\begin{eqnarray}
 \Gamma^{\rm exp}(D^{*+})\left[1-{\mathcal B}(D^{*+}\to D^+ \gamma)\right]&=&\Gamma(D^{*+}\to D^0 \pi^+)+\Gamma(D^{*+}\to D^+ \pi^0)\, \nonumber\\
 &=& \frac{2\,m_{D^0}\,|\vec{k}_{\pi^+}|^3+m_{D^+}\,|\vec{k}_{\pi^0}|^3}{12\, \pi\,m_{D^{*+}}\,f_{\pi}^2}\,\hat{g}^2\,,
\end{eqnarray}
where $|\vec{k}_{\pi^+}|=\frac{\sqrt{[m_{D^*}^2-(m_D+m_{\pi})^2]\,[m_{D^*}^2-(m_D-m_{\pi})^2]}}{2\,m_{D^*}}$ is the three-momentum of pion in the rest frame of $D^*$ meson. Using the inputs listed in Table~\ref{tab:Inputs}, we get numerically
\begin{equation}
\hat g=0.61 \pm 0.07,
\end{equation}
which is a bit larger than both the LQCD simulation and our results. This discrepancy might be due to the fact that the charm quark is not very heavy and there are potentially large ${\cal O}(1/m_c^n)$ corrections to the relation Eq.~(\ref{eq:ghat-def}) with $B$ replaced by $D$.

\section{Conclusions}

In this paper, motivated by the precisely measured $q^2$ spectrum of semi-leptonic $B\to \pi \ell \nu$ decays by the BaBar~\cite{:2010uj,delAmoSanchez:2010zd,Aubert:2006px} and Belle~\cite{Ha:2010rf,Hokuue:2006nr} collaborations, we have performed a phenomenological study of the strong coupling $g_{B^* B\pi}$ and the normalized coupling $\hat g$ appearing in the HMChPT, which is related to the pole residue of the vector form factor $f_+(q^2)$ at the unphysical point $q^2=m_{B^*}^2$.

Through a detailed analysis, we found that the extracted values based on the BK, BZ and BCL parametrizations are consistent with each other and also roughly in agreement with other theoretical and lattice estimates, while the BGL ansatz gives much larger values, which is due to the spurious zero at $q^2=t_+$ in definition of the outer function $\phi(q^2,q^0)$. It is also found that the errors increase with more expansion parameters added in the BGL and BCL parametrizations, leading to a loss of predictive power; the BGL and BCL parametrizations with more fitting parameters could not be well constrained by the current data in the physical region.

In order to gain further information about the $q^2$ behavior of heavy-to-light transition form factors, much more precise experimental data on exclusive semi-leptonic B-meson decays, as well as additional  information on the behavior of the vector form factor $f_+(q^2)$ outside the physical region are urgently needed.

\section*{Acknowledgments}

The work was supported in part by the National Natural Science Foundation under contract Nos.~11075059, 10735080, 11005032 and 11047165. X.~Q. Li was also supported in part by MEC (Spain) under Grant FPA2007-60323 and by the Spanish Consolider Ingenio 2010 Programme CPAN (CSD2007-00042).

\end{document}